\documentclass[a4paper,fleqn,usenatbib]{mnras}

\usepackage[T1]{fontenc}
\usepackage{ae,aecompl}
\usepackage[dvipsnames]{xcolor}

\usepackage{graphicx}	
\usepackage{epstopdf}
\usepackage{amsmath}	
\usepackage{amssymb}	
\usepackage{listings}

\newcommand{\msun}{$h^{-1}$~M$_\odot$}

\title[The Lomonosov simulations]{Pushing down the low-mass halo concentration frontier with the Lomonosov cosmological simulations}

\author[S. V. Pilipenko et al.]{
Sergey V. Pilipenko,$^{1}$\thanks{spilipenko@asc.rssi.ru (SVP)}
Miguel A. S\'anchez-Conde$^{2,3,4}$\thanks{sanchezconde@fysik.su.se (MASC)}, 
Francisco Prada$^{5}$, 
\newauthor
Gustavo Yepes$^{4}$
\\
$^{1}$Astro Space Center, P. N. Lebedev Physical Institute of RAS, Profsojuznaya 84/32, Moscow 117997, Russia\\
$^{2}$ Oskar Klein Centre for Cosmoparticle Physics, Department of Physics, Stockholm University, SE-10691 Stockholm, Sweden \\
$^{3}$ Instituto de F\'{\i}sica Te\'orica UAM/CSIC, Universidad Aut\'onoma de Madrid, E-28049 Madrid, Spain\\
$^{4}$ Departamento de F\'isica Te\'orica, M-15, Universidad Aut\'onoma de Madrid, E-28049 Madrid, Spain\\
$^{5}$ Instituto de Astrof\'{\i}sica de Andaluc\'{\i}a (IAA-CSIC), Glorieta de la Astronom\'{\i}a, E-18008 Granada, Spain\\
}

\date{Accepted XXX. Received YYY; in original form ZZZ}

\pubyear{2017}

\begin{document}
\label{firstpage}
\pagerange{\pageref{firstpage}--\pageref{lastpage}}
\maketitle

\begin{abstract}
We introduce the Lomonosov suite of high-resolution N-body cosmological simulations covering a full box of size 32 $h^{-1}$~Mpc with low-mass resolution particles ($2\times10^7$ \msun) and three zoom-in simulations of overdense, underdense and mean density regions at much higher particle resolution ($4\times10^4$ \msun). 
The main purpose of this simulation suite is to extend the concentration-mass relation of dark matter halos down to masses below those typically available in large cosmological simulations. 
The three different density regions available at higher resolution provide a better understanding of the effect of the local environment on halo concentration, known to be potentially important for small simulation boxes and small halo masses. 
Yet, we find the correction to be small in comparison with the scatter of halo concentrations. We conclude that zoom simulations, despite their limited representativity of the volume of the Universe, can be effectively used for the measurement of halo concentrations at least at the halo masses probed by our simulations.
In any case, after a precise characterization of this effect, we develop a robust technique to extrapolate the concentration values found in zoom simulations to larger volumes with greater accuracy.
All together, Lomonosov provides a measure of the concentration-mass relation in the halo mass range  $10^7-10^{10}$ \msun~with superb halo statistics. 
This work represents a first important step to measure halo concentrations at intermediate, yet vastly unexplored halo mass scales, down to the smallest ones.  
All Lomonosov data and files are public for community's use.
\end{abstract}

\begin{keywords}
cosmology-simulations --- large scale structure --- dark matter halos
\end{keywords}


\section{Introduction}

According to the standard $\Lambda$CDM cosmological framework, the Universe is populated with dark matter halos which host the most of light-emitting material (stars, hot gas). 
The mass spectrum of cold dark matter halos spans over many orders of magnitude, from the largest virialized structures observed in the Universe with masses above $10^{15} {\rm M}_\odot$, extending down to very small masses, probably as small as the Earth's, i.e. $10^{-6} {\rm M}_\odot$, or even less \citep{2004MNRAS.353L..23G,2006PhRvL..97c1301P,Diamanti:2015kma}.\footnote{The precise value of this minimum halo mass is set by processes that depend on the chosen particle physics and cosmological models, such as free streaming of dark matter particles or the effect of acoustic oscillations (see, e.g.,~\citet{Zybin:1999ic,Loeb:2005pm}), with possible values ranging within $10^{-12} - 10^{-4} {\rm M}_\odot$.}  
 
 Halos lighter than $\sim10^{7} {\rm M}_\odot$~are most probably not able to host stars or gas at all, thus remaining completely dark~\citep{1992MNRAS.256P..43E,2000ApJ...539..517B,2000ApJ...542..535G,2013PhR...531....1S}. Yet, there is a hope to detect them via gravitational lensing~\citep{1998MNRAS.295..587M,2001ApJ...563....9M,2002ApJ...565...17C,2009MNRAS.398.1235X,2016PhRvD..94d3505C}, dark matter particle decay or annihilation~\citep{Baltz:1999ra, 2004PhRvD..69d3501K,2005Natur.433..389D, 2007ApJ...659L.125B,2008MNRAS.384.1627P,Ackermann:2012nb, Mirabal:2012em, Zechlin:2012by, Berlin:2013dva, Bertoni:2015mla, Schoonenberg:2016aml,2016JCAP...09..047H} or distortions in stellar velocity distribution \citep{2012ApJ...760...75C, 2015MNRAS.446.1000F, 2015ApJ...808...15C,2016ApJ...819....1I}. Low-mass halos may also have an important impact on future observations of reionization processes~\citep{2002ApJ...572L.123I,2005SSRv..116..625C}. Therefore, a precise characterization of the statistical and structural properties of low-mass halos, such as their mass spectrum or density profiles, is needed in order to make accurate predictions for future studies and to drive their search. Also, a better knowledge of the low-mass halo population will potentially help in refining the estimates for the so-called substructure boost factor to dark matter annihilation signals, whose exact value can largely impact the identification of the best astrophysical targets for dark matter searches \citep{Bergstrom:1998jj, Stoehr:2003hf, Lavalle:1900wn, Kuhlen:2008aw, 2013JCAP...04..009A, Sanchez-Conde:2013yxa, 2014ApJ...788...27I, Zavala:2015ura,2017MNRAS.tmp...47M}. Finally, by studying the smallest halo mass scales we will be testing fundamental aspects of the underlying cosmological model.

Dark matter halos in simulations are usually characterized by a spherical density profile, fitted by a Navarro-Frenk-White (NFW) formula~\citep{1997ApJ...490..493N}. This description of the density distribution has two parameters, i.e., halo virial mass and concentration parameter. The latter is defined as $c\equiv R_\mathrm{vir}/r_s$, where $r_s$ is a scale radius that, in the case of NFW, sets the transition from $r^{-1}$ to $r^{-3}$ in the profile. Other fits different from NFW are often used to describe density profiles of halos as well, e.g. the Einasto profile~\citep{1965TrAlm...5...87E,2004MNRAS.349.1039N,2008MNRAS.387..536G,2014MNRAS.441.3359D}.\footnote{However, it is also possible to calculate halo concentrations in a way which does not depend on the particular choice made for the density profile~\citep{2008Natur.454..735D,klypin,Klypin:2014kpa,2017MNRAS.tmp...47M}.}
Analysis of parameters of millions of halos has shown that, for a fixed mass and redshift, halos populate only a certain range of concentrations. A dependency of halo median concentration on mass and redshift is found~\citep{1997ApJ...490..493N,2001MNRAS.321..559B,2001ApJ...554..114E,2002ApJ...568...52W,2003MNRAS.339...12Z,2007MNRAS.381.1450N,2008MNRAS.387..536G,2009ApJ...707..354Z,2011MNRAS.411..584M,klypin,Prada:2011jf,2014MNRAS.441..378L,2015ApJ...799..108D,Klypin:2014kpa}. 

The behavior of concentrations for intermediate-mass halos of around $10^{9}$ -- $10^{14} {\rm M}_\odot$ is well known, while analyzing concentrations at lower and higher masses remains a computational challenge. At the highest mass end, halos are very rare (the mass function is nearly exponential), thus large simulation boxes are needed while also having high enough resolution to measure concentrations (see, however, e.g.,~\citet{Klypin:2014kpa} for recent results).
Simulating low-mass ($<10^9$ \msun) halos using a single test particle mass across the simulation box results in a small box size that becomes much smaller than the typical homogeneity scale of the large-scale structure of the Universe, $\sim$100 $h^{-1}$~Mpc. Also, at low redshifts, nonlinear effects come to play in a small box~\citep{2005MNRAS.358.1076B}. Zoom-in simulations are used to partially solve this problem. They do not suffer from non-linearity; yet, the issue of ``representativity'' in the simulated volume remains. 

The concentration of halos is expected to depend on the local environment from theory using simple considerations (e.g.,~\citet{2002ApJ...568...52W}), as the concentration reflects the density of the Universe at the time and place of halo formation. This effect was observed to be quite weak in the early simulations using small boxes to access low mass halos (e.g.,~\citet{2007MNRAS.378...55M}), but is found to be significant in more recent works (e.g.,~\citet{2017MNRAS.466.3834L}). This means that neglecting the representativity problem in the study of halo concentrations at small mass scales is not possible.


The problem of representativity has been solved in the past, e.g., by computing the number density fluctuations of minihalos at the epoch of reionization~\citep{2004ApJ...609..474B, 2007ApJ...654...12Z, 2008ApJ...689....1S,2009ApJ...703L.167A, 2011MNRAS.411..955M, 2015MNRAS.450.1486A}.
The solution involved, first, a quantification of the so-called halo bias (dependency of halo number density on the local environment density) and, then, the generation of a coarse large-scale density distribution which was later filled with halos according to the earlier found bias. In this paper, we will use a similar approach to extrapolate halo concentrations found in small simulated volumes to larger ones.

Current progress in establishing the concentration-mass relation for low-mass halos can be summarized as follows. There are a few simulations available of the first halos that were formed in the early Universe, i.e those with masses just above the  low-mass cut-off in the halo mass function. In particular, the latter was set to $\sim 10^{-6} {\rm M}_\odot$ in all these simulations, being the typical masses of the microhalos  $\sim 10^{-6}-10^{-3} {\rm M}_\odot$ ~\citep{2005Natur.433..389D,2013JCAP...04..009A,2014ApJ...788...27I}. The simulations were performed in tiny boxes of size 30--400 pc and they all stop at very early redshifts, $z\sim30$. 
In addition, already at much larger halo masses of the order of $10^{7}-10^{9} {\rm M}_\odot$, a few halo concentration values were obtained by \citet{Sanchez-Conde:2013yxa} as a by-product from high resolution zoom-in simulations of a Milky Way-like object~\citep{2008Natur.454..735D}. Finally, there are also a few high-resolution simulations of dwarf-like objects at slightly larger masses but still below $10^{10} {\rm M}_\odot$~\citep{2004ApJ...612...50C,Ishiyama:2011af,2016MNRAS.457.3492H}. In short, there are no halo concentration results yet in the range $\sim 10^{-3}-10^{7}  {\rm M}_\odot$~while for $10^{7}-10^{9} {\rm M}_\odot$~the statistics is not very representative. See \citet{Sanchez-Conde:2013yxa} for a full discussion of current results.

The purpose of this paper is to start filling the aforementioned gap in the results on halo concentrations, $\sim 10^{-3}-10^{7} {\rm M}_\odot$. As a first step, we populate the $10^7-10^9 {\rm M}_\odot$~halo mass range with much better statistics than currently available using the Lomonosov new suite of cosmological simulations. We do so by performing new several zoom-in simulations representative of different kinds of local environments. We complement them with data from the Small MultiDark Planck cosmological simulation (SMDP) at larger halo masses \citep{Klypin:2014kpa}. Building upon that gained from the analysis of these simulations, we then develop a robust method of extrapolating the zoom-in concentration results to the full simulation volume with enough confidence. 

We share our results with the community by making the most relevant simulation data of this study publicly available. This includes simulation snapshots, halo catalogs and concentration values. We refer the interested readers to the \href{http://projects.ift.uam-csic.es/skies-universes/SUwebsite/index.html}{\it Skies and Universes} website\footnote{http://projects.ift.uam-csic.es/skies-universes/SUwebsite/index.html} in order to access this material. 

The paper is organized as follows. We describe the new set of Lomonosov simulations in section \ref{sec:simulations}, providing full technical details of how they were ran and analyzed. In section \ref{sec:analysis} we perform the measurement of halo concentrations in the new simulations and study its dependence with the local environment. We analyze data from Small MultiDark Planck box in this section too, which provides us with additional clues on this effect. We then correct the Lomonosov zoom-in halo concentration results by the effects of environment by proposing and applying our extrapolation technique in section \ref{sec:context}, and put the Lomonosov concentration values into a more general context in this same section. We conclude with summarizing our main findings in section \ref{sec:conclusions}.

\section{The Lomonosov Simulations and halo selection}  \label{sec:simulations}
The Lomonosov suite consists of four dark-matter-only simulations run at the Lomonosov supercomputer of the Moscow State University computer center.\footnote{http://parallel.ru} The box size of each simulation is 32 $h^{-1}$~Mpc, all sharing the same Planck cosmological parameters \citep{planck}, i.e. $\Omega_\Lambda=0.692885$, $\Omega_m=0.307115$, $h=0.6777$, $\sigma_8=0.8288$, $n_s=0.9611$. The initial conditions were prepared using the massively-parallel zoom \texttt{ginnungagap} public code.\footnote{https://github.com/ginnungagapgroup/ginnungagap} The simulations were then run using the public version of \texttt{GADGET-2}\footnote{http://wwwmpa.mpa-garching.mpg.de/~volker/gadget/} \citep{springel} starting from redshift $z_{init}=99$ and adopting a total matter (DM+baryons) initial power spectrum from \texttt{CAMB}\footnote{http://camb.info/}.

\begin{table}
\caption{Most relevant parameters of the Lomonosov simulation suite. $(x,y,z)$ are the zoom region center coordinates in $h^{-1}$~Mpc, $R$ is the radius of the zoom region at $z=1$. The overdensity in the last row has been defined with respect to the mean density of the \texttt{L512} low-resolution box.} 
\centering
 \begin{tabular}{lcccc}
  \hline
\hline
  Parameter & \texttt{L512} & \texttt{L4km} & \texttt{L4ko} & \texttt{L4ku} \\ \hline
  Box size ($h^{-1}$~Mpc) & 32 & 32 & 32 & 32 \\
  No. of particles & $512^3$ & $4096^3$ & $4096^3$ & $4096^3$ \\ 
  Mass res. (\msun) & $2\times10^7$ & $4\times 10^4$  & $4\times 10^4$  & $4\times 10^4$  \\ 
  Force res. (kpc/h) & 1.5 & 0.2 & 0.2 & 0.2 \\ 
  Initial redshift & 99 & 99 & 99 & 99 \\ 
  $x,y,z$ ($h^{-1}$~Mpc) & - & 7,18,10 & 12,23,11 & 22,12,17 \\ 
  $R$ ($h^{-1}$~Mpc) & - & 3 & 3 & 5 \\ 
  Overdensity & 1.0 & 1.02 & 2.15 & 0.37 \\ 
 \hline
   \hline
 \end{tabular}
\label{tab:sim}
\end{table}

A first, low-resolution simulation was run for the full box with $512^3$ particles (\texttt{L512}). The three other high-resolution simulations correspond to three different zoom regions of this box with an effective resolution of $4096^3$ particles. This gives a mass resolution of $4\times 10^4$ \msun. The size of each of the zoom regions was limited by computer time to about $5\times10^8$~particles. The regions were chosen to represent three different types of large-scale environment: a region with approximately the mean density (\texttt{L4km}), a void (\texttt{L4ku}) and an overdense region (\texttt{L4ko}).

The most relevant parameters of the four simulations and corresponding zoom regions are given in Table \ref{tab:sim}. Simulation snapshots, halo catalogs and concentration values as well as images and video visualization of the simulations can be found at the \href{http://projects.ift.uam-csic.es/skies-universes/SUwebsite/index.html}{\it Skies and Universes} website.\footnote{http://projects.ift.uam-csic.es/skies-universes/SUwebsite/index.html} Simulation snapshots were produced at redshifts 1, 1.22, 2, 3, 4, 5, 7, 9, 15, 19. The minimal redshift of the simulations was set to one in order to reduce artefacts due to box periodicity.

We use the \texttt{Rockstar} halo finder~\mbox{\citep{Behroozi:2011ju}} to select halos, and define their masses, $M_{200}$, as those corresponding to spherical overdensities equal to $200$ times the critical density, including all particles (bound plus unbound). The corresponding radius enclosing $M_{200}$ is the halo virial radius, $R_{200}$.
Inside the zoom regions, we select only those halos which are located at least 0.5 $h^{-1}$~Mpc far from the boundary of the region, and check that halos do not contain high-mass (i.e., low-resolution) particles.
We refer the reader to Appendix \ref{ax:rockstar} for details on the \texttt{Rockstar} setup.

\section{Measurement of Lomonosov halo concentrations}   \label{sec:analysis}
For the analysis of the concentration of halos in the Lomonosov simulations, we select all distinct halos with more than 250 particles\footnote{This number represents a good compromise, as it allows us to study the smallest halo masses that are possible to probe with our simulations, at the same time ensuring a decent description of the internal structure of halos (and, thus, a good measure of their concentrations).} and adopt the method described in \citet{klypin,Prada:2011jf} to measure concentrations. To do so, we measure $V_\mathrm{max}$, the maximal circular velocity, and $V_\mathrm{200}$, the circular velocity at the virial radius. Then, we solve the following equation for each halo to find the concentration, $c_{200}$:
\begin{equation}
\left( {V_\mathrm{max} \over V_\mathrm{200}} \right)^2 = {0.216~c_{200} \over f(c_{200})}, \;\; f(c) = \ln(1+c)-{c\over 1+c}.
\end{equation}
Note that the equation (1) implicitly adopts an NFW density profile for the Lomonosov halos. Furthermore, we use all distinct halos found in the simulations, i.e. we do not apply any relaxation criteria or take subhalos.

\begin{figure}
    \centering
    \includegraphics[width=\linewidth]{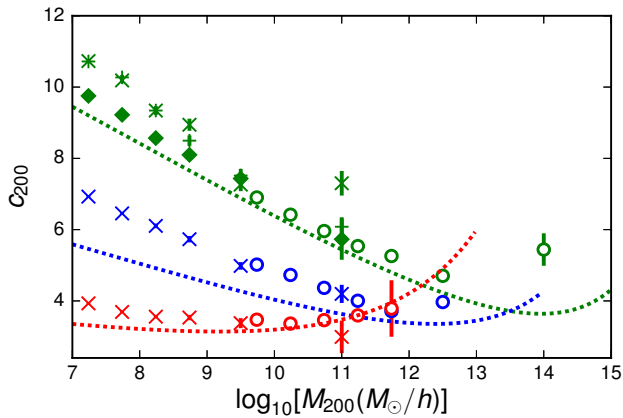}
    \caption{Median concentrations and 1$\sigma$ statistical errors of Lomonosov halos at $z=1$ (green), $z=2$ (blue) and $z=5$ (red) as a function of mass for the four simulation runs: \texttt{L512} (open circles), \texttt{L4km} (crosses), \texttt{L4ku} (plus signs) and \texttt{L4ko} (diamonds). Dotted lines correspond to the $c(M)$ fits obtained from the MultiDark simulations for those redshifts (\citet{Klypin:2014kpa}, eq. (25)).}
    \label{fig:call}
\end{figure}

First, we check for consistency of our simulation results with those found in previous works. The concentration-mass relation, $c(M)$, for the Lomonosov simulations at $z=1,2,5$ is shown in Fig. \ref{fig:call}. We find a systematic $\sim$5-10\% disagreement with the concentration results found in the suite of MultiDark cosmological simulations at larger masses \citep{Klypin:2014kpa}. We note that this level of disagreement is nevertheless comparable to the precision of the fit to MultiDark $c(M)$ data (see \citet{Klypin:2014kpa}).

Most notably, from Fig. \ref{fig:call} it can be seen that there is a significant difference in concentrations between \texttt{L4ko}, \texttt{L4km} and \texttt{L4ku} simulations at $z=1$, i.e., $c(M)$ is indeed sensitive to the density of the environment (see also, e.g., Fig. 5 in \citet{2017MNRAS.466.3834L}). Due to the way these zoom regions were selected, with overdensities of 0.37, 1.02 and 2.15 times the mean density of the \texttt{L512} box, respectively, one could expect the median $c(M)$ averaged over the full box to reside somewhere in between the concentration results for \texttt{L4ko} and \texttt{L4ku}, and, probably, close to \texttt{L4km}. This effect will need to be quantified in detail, as one of our goals is to explore the $c(M)$ relation down to very small masses -- since one cannot simulate the full box -- and we find that different zoom regions exhibit different $c(M)$ relations. Only then it will be possible to extrapolate halo concentrations as measured in zoom regions to larger volumes with enough confidence. 

\begin{figure}
    \centering
    \includegraphics[width=\linewidth]{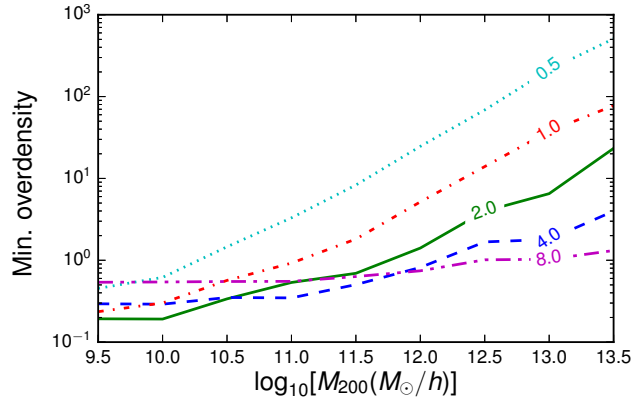}
    \caption{Minimum cell overdensity in which a halo could be found vs. its mass for different cell sizes: 8 $h^{-1}$~Mpc (pink), 4 $h^{-1}$~Mpc (blue), 2 $h^{-1}$~Mpc (green), 1 $h^{-1}$~Mpc (red), 0.5 $h^{-1}$~Mpc (cyan).}
    \label{fig:mindens}
\end{figure}

In order to characterize the large-scale density field, we compute the density in cells using the cloud-in-cell (CiC) approach from the particle data of the \texttt{L512} full box simulation. We then extract the density at each halo position using the same CiC approach. We use cells of size 8, 4, 2 and 1 $h^{-1}$~Mpc. It must be noted that the existence of a massive halo inside a cell restricts the minimal density that this can have. Indeed, for large halos, the CiC method does not retrieve the overdensity of the environment but the overdensity of the halo itself. On the other hand, the larger is the cell size, the less are the deviations from the mean density. This is illustrated in Figure \ref{fig:mindens}. We use 2 $h^{-1}$~Mpc cells for the subsequent analyses as a good compromise to maximize the range of overdensities that we are going to probe for halos with masses $\lesssim10^{11}$ \msun. It can be seen in Figure \ref{fig:mindens} that for cell sizes of 0.5 or 1 $h^{-1}$~Mpc the minimal overdensity is $\geq 1.0$ for $\sim10^{11}$ \msun~halos, and, while for cell sizes of 4 or 8 $h^{-1}$~Mpc the curves are much flatter, at $\sim10^{10}$ \msun~the minimal overdensity is higher than in the case of 2 $h^{-1}$~Mpc cells. 


\begin{figure*}
    \centering
     \includegraphics[width=0.49\textwidth]{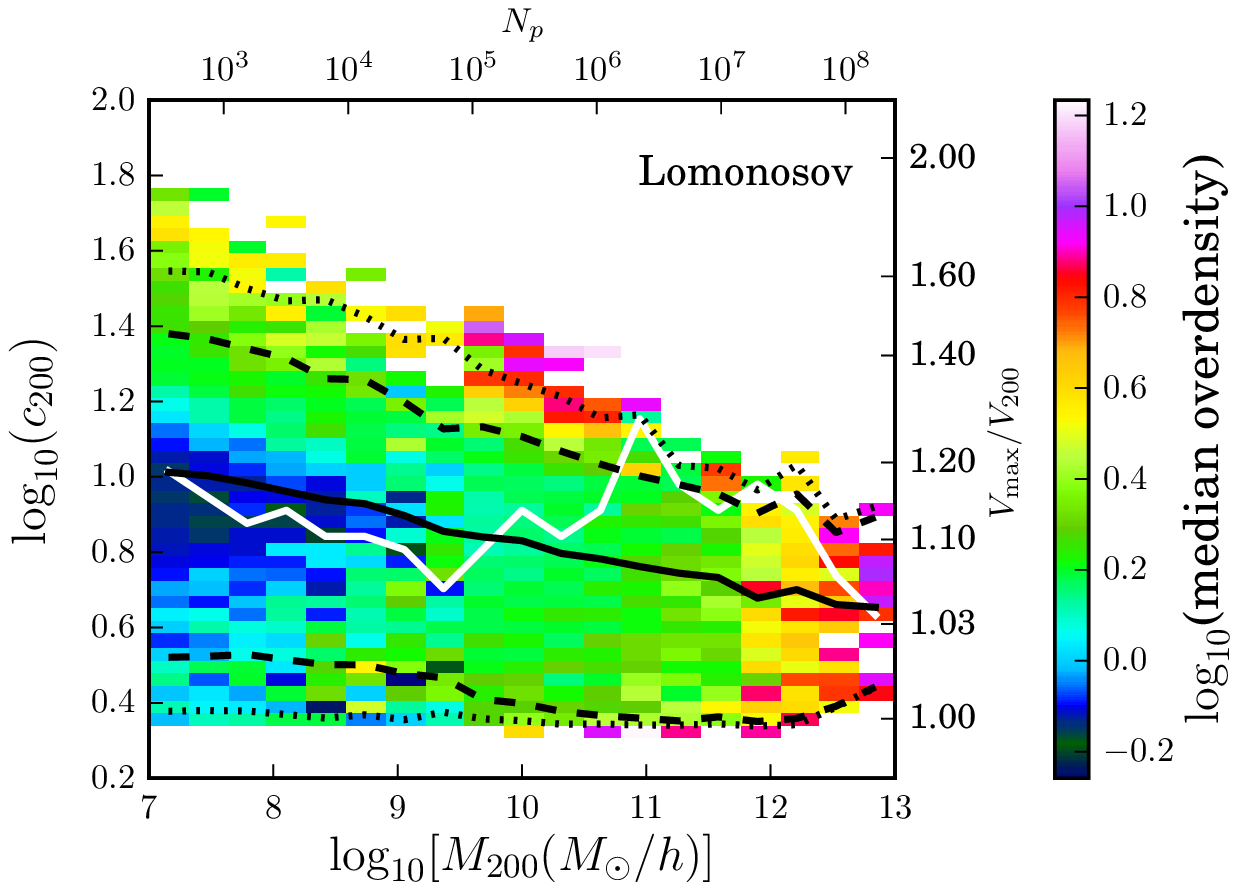}
     \includegraphics[width=0.49\textwidth]{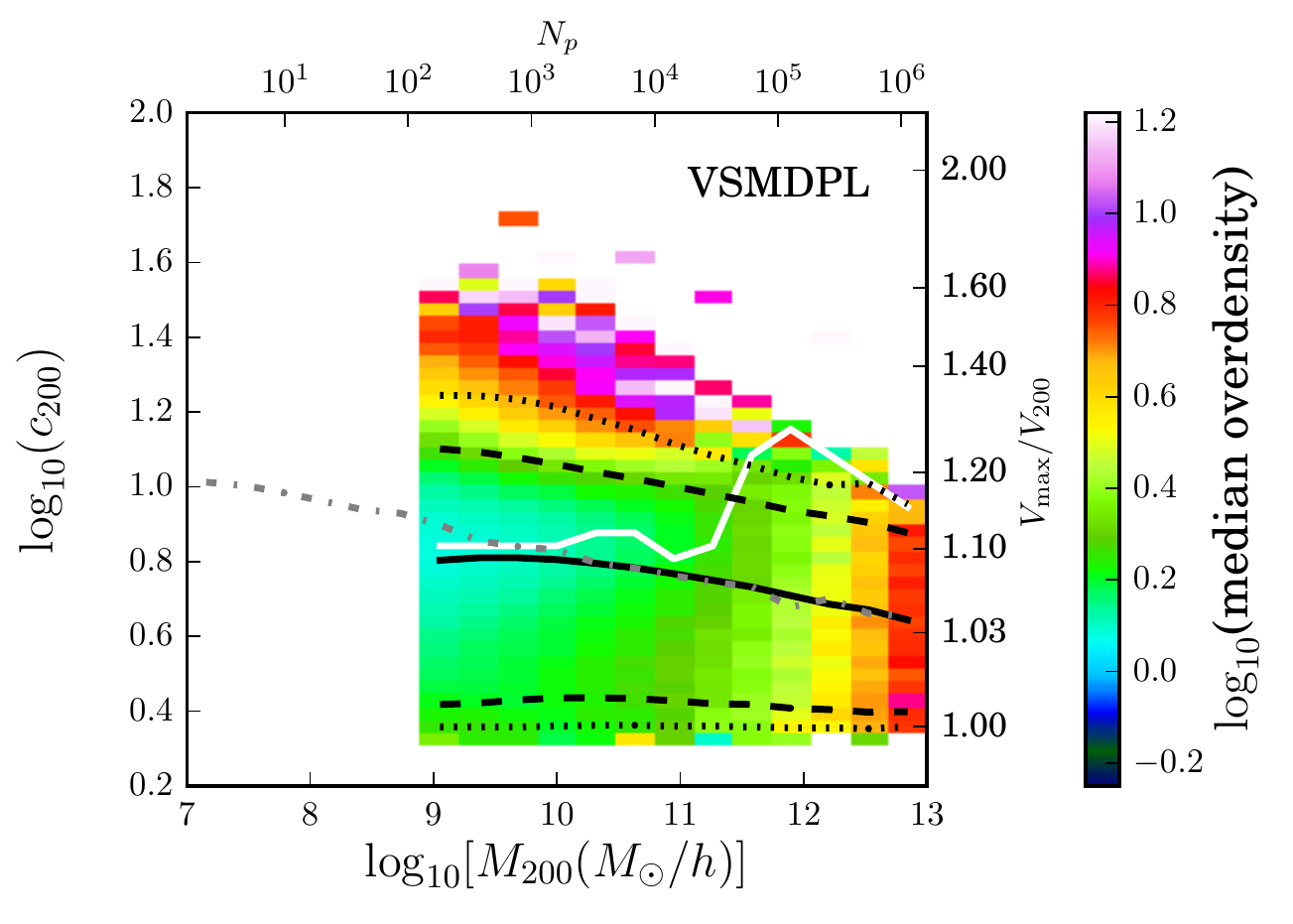}\\
     \includegraphics[width=0.49\textwidth]{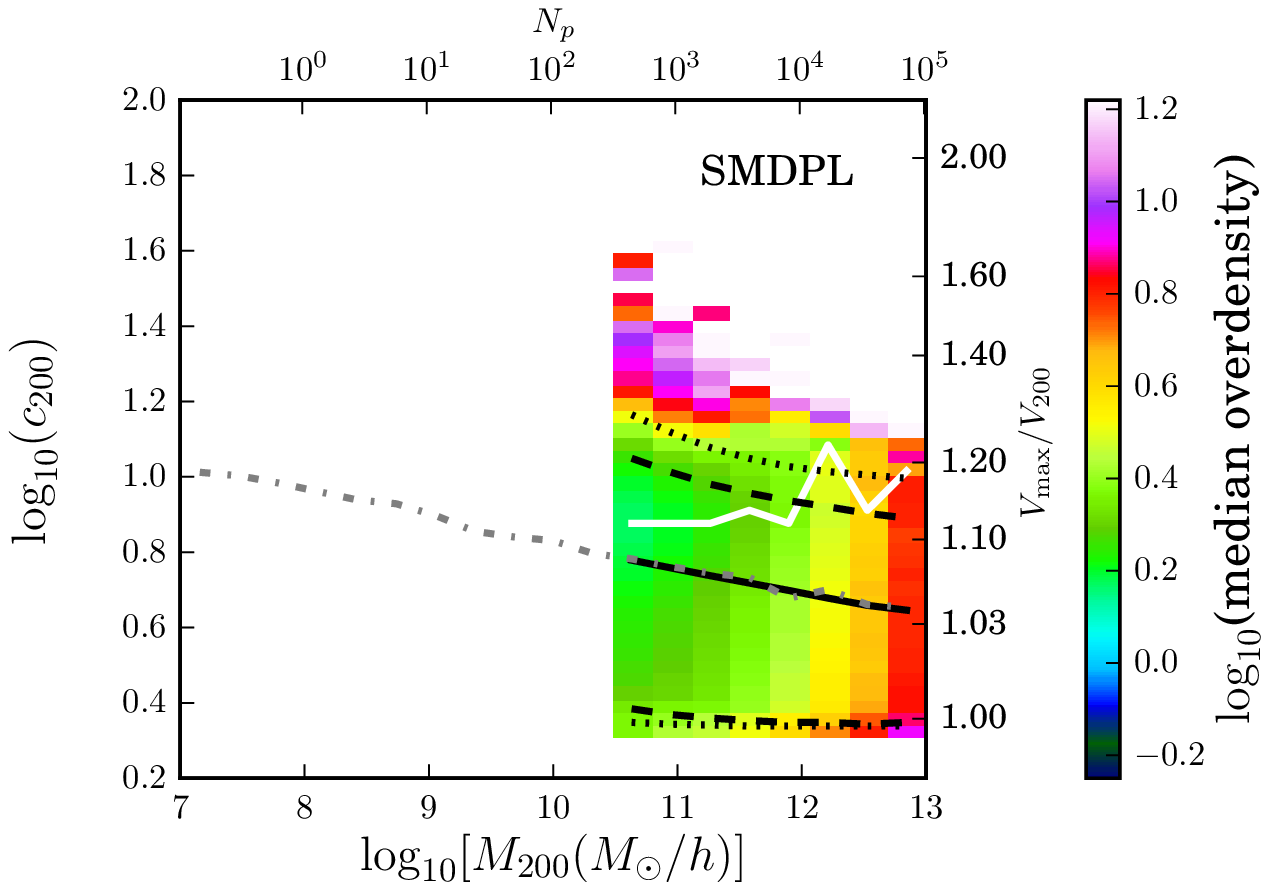}
     \includegraphics[width=0.49\textwidth]{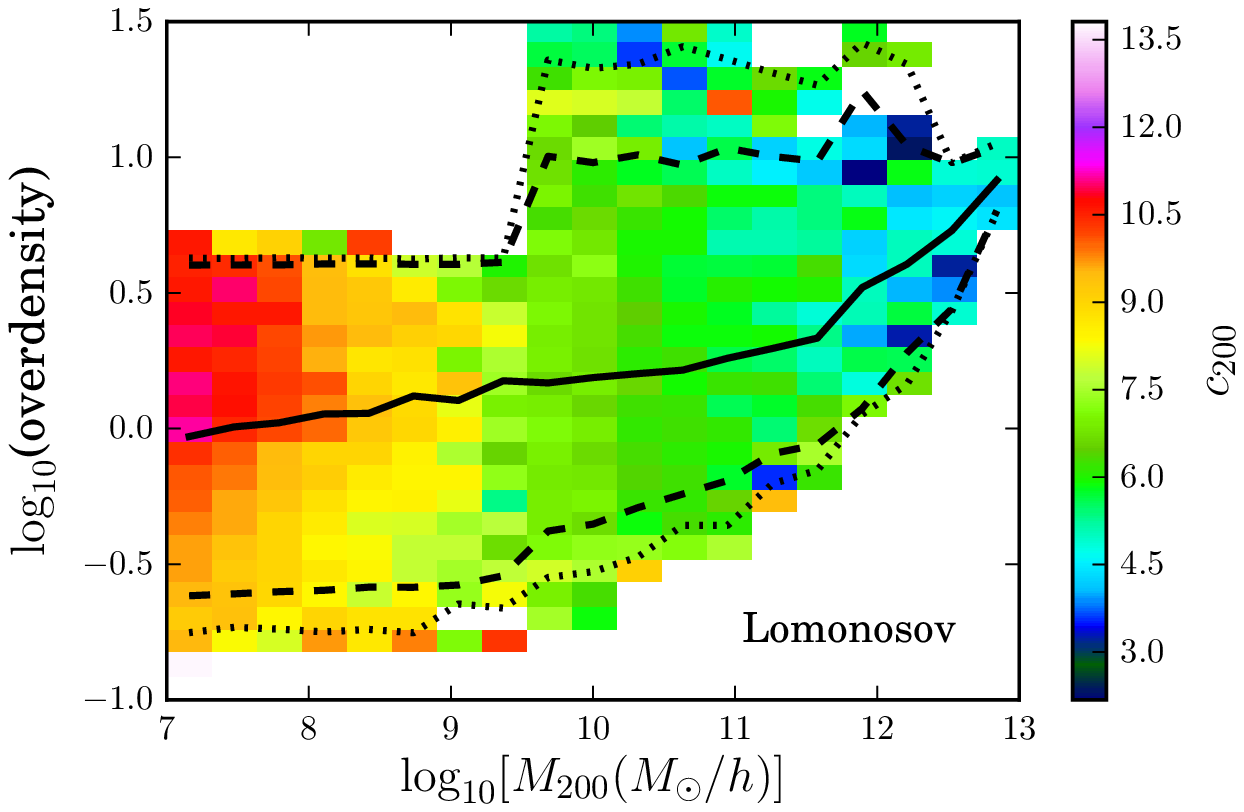}
   \caption{{\it Top panels and bottom left:} Median halo large-scale overdensities as a function of mass and concentration for the full Lomonosov suite (top left), Very Small MultiDark Planck (top right) and Small MultiDark Planck (bottom left). All results are for $z=1$. The black solid lines represent median concentration values. Dashed and dotted lines represent, respectiverly, 2 and 3$\sigma$ scatter. The abrupt transition at $M\sim 5\times10^{9.5}$ \msun~ in the case of Lomonosov panels corresponds to the minimal mass of halos in the \texttt{L512} simulation, which possesses a different volume selection with respect to the zoom simulations and therefore same mass halos inhabit different environments in comparison. The dot-dashed line on the top right and bottom left panels represents the median concentration from the left panel. The white line shows the local minimum of median overdensity.
{\it Bottom right panel:} Median halo concentrations as a function of mass and large-scale overdensity as measured in 2 $h^{-1}$~Mpc size cells for the full Lomonosov suite. Note that the zoom regions cover a limited range of overdensities which is seen as an abrupt change at $M\sim3\times10^9$ \msun.}
    \label{fig:cmd}
\end{figure*}

Next, we group all halos in the full Lomonosov suite (i.e., the four simulations) into 30 equal logarithmic bins of $M_{200}$ between $10^7$ and $10^{13}$ \msun~and 50 logarithmic bins of concentration between 2 and 80.
The obtained concentration-mass-density relation for Lomonosov halos is depicted in the left panel of Fig. \ref{fig:cmd}. This figure shows that at low halo masses, for which there is a proper coverage of overdensity values in our simulations,\footnote{See Appendix \ref{ax:extrapol} for further details on this statement.} less concentrated halos inhabit less dense regions and vice-versa, in good agreement with what it has been found in previous work \citep{2017MNRAS.466.3834L} but confirmed now at even lower halo masses thanks to Lomonosov data. Also, halos with typical concentrations (i.e., close to the median) seem to populate the least dense regions at a fixed mass. At the same time, as expected, the environment density for typical halos is observed to grow with halo mass. 

In order to test the effect of numerical resolution on the measurements of concentration dependence on mass and environment overdensity, we repeat the same analysis for two simulations: the SMDPL simulation \citep{Klypin:2014kpa} and VSMDPL\footnote{Available at www.cosmosim.org.}. SMDPL (Small MultiDark Planck) has a box size of 400 $h^{-1}$~Mpc and a mass resolution comparable to that of \texttt{L512} ($10^8$ \msun). VSMDPL (Very Small MultiDark Planck) has a box size of 160 $h^{-1}$~Mpc and particle mass $6.2\times 10^6$ \msun. Both simulations use the same cosmological parameters as Lomonosov. As before, we use CiC density assignment with cell sizes of 1, 2 and 4 $h^{-1}$~Mpc to characterize the density field. The results for both VSMDPL and SMDPL are shown in Fig. \ref{fig:cmd}.
The analysis of these results gives us more clues on the dependency of halo concentration on the environment, and allows us to confirm with greater confidence our findings in Lomonosov data discussed above. 


In particular, one can see from the top right panel of Fig. \ref{fig:cmd} (VSMDPL) that, for halos with masses below $\sim 3\times10^{11}$ \msun, the most highly concentrated halos (which amounts to 3\% of all halos) typically inhabit overdense regions. Halos with median concentrations and below inhabit regions with median overdensity close to the mean density of the Universe. Interestingly, the median overdensity for halos with concentrations below the median starts to slowly increase when the concentration decreases. To make this more clear, we plot in Fig. \ref{fig:cmd} the position of the local minimum of overdensity (white lines). For masses below $\sim 3\times10^{11}$ \msun~this minimum roughly coincides with the median concentration.  This demonstrates that the concentration of low mass halos is connected with overdensity in a complex way, and the environmental effects are important for measuring these concentrations.
The comparison of the results of SMDPL and VSMDPL simulations in the same mass range but with different numerical resolution allows us to conclude that for the halos with more than 1000 particles the described above trends are real and not mere artefacts.


\section{Halo concentration-mass relation}  \label{sec:context}

Our study of the impact of the environment on the concentration-mass relation in the zoom simulations allows us to reliably extrapolate the zoom simulation results to the full box by applying an appropriate correction. We follow the next steps:
\begin{enumerate}
\item We define four mass bins with edges $[10^7,3\times10^7,10^8,10^9,10^{10}]$ \msun.
\item For each mass bin we make a 2D histogram of halo overdensities (as defined in Section \ref{sec:analysis}) and logarithm of concentrations. 
\item We multiply the number of haloes in each overdensity bin by a {\it volume correction factor}, defined as below, and sum the result over all density bins to achieve the corrected 1D distribution of concentrations. 
\item We find the median of this binned concentration distribution by making a cumulative sum of it and using a linear interpolation to find the 50 percentile level.
\end{enumerate}
The volume correction factor is computed in the following way. 
We place test particles on a uniform grid with cell size of 0.32 $h^{-1}$~Mpc. We measure CiC densities on the coarser 2 $h^{-1}$~Mpc cell-size grid described in Section \ref{sec:analysis} for each test particle and group them into bins of density. The volume correction factor for a given density bin is the ratio between the number of test particles in this bin belonging to the full box and the number of those belonging only to the zoom regions. We test our method of extrapolation in several ways, and refer the interested reader to Appendix \ref{ax:extrapol} for results on our validation tests. 

The corrected concentrations are presented in Table \ref{tab:corr} together with the uncorrected concentrations  originally measured in each zoom simulation volume.\footnote{We provide corrected median $V_{max}/V_{200}$ values and 1$\sigma$ scatter in Table \ref{tab:vmax} of Appendix \ref{ax:vmax}.} The mean corrections we found with respect to the measured concentrations in the $10^7-10^9$ \msun~halo mass range are $-0.012$, $-0.011$ and $0.03$ for the mean density simulation, the overdense region, and the underdense region, respectively. We note that in all cases the correction is small in comparison with the scatter of halo concentrations. This allows us to conclude that zoom simulations, despite their limited representativity of the volume of the Universe, can be effectively used for the measurement of halo concentrations at low halo masses.

\begin{figure*}
    \centering
    \includegraphics[width=0.88\linewidth]{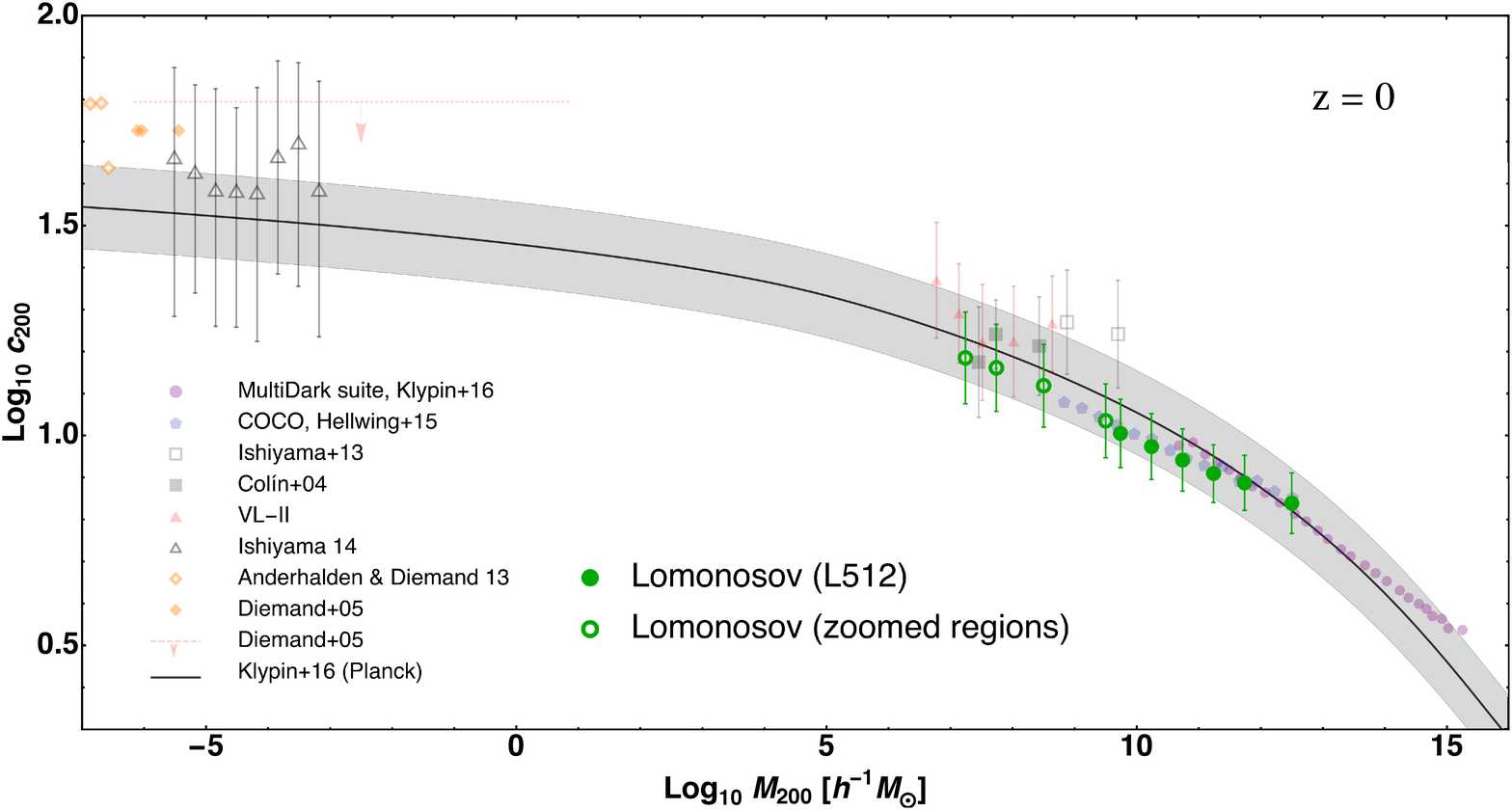}
    \caption{Lomosonov median concentration values (green filled circles for \texttt{L512}, open circles for zoomed regions), with  $1\sigma$-error bars, in comparison with other simulation data sets at different halo mass scales \citep{2004ApJ...612...50C,2005Natur.433..389D,Diemand:2008in,Ishiyama:2011af,Anderhalden:2013wd,Sanchez-Conde:2013yxa,Ishiyama:2014uoa,2016MNRAS.457.3492H,Klypin:2014kpa}; see legend for specific symbols. All concentration values but those of the MultiDark suite (purple circles without error bars) and VL-II (red triangles) were extrapolated down to $z=0$ by applying the corresponding $[H(z)/H(0)]^{2/3}$ correction factor; see text for details.  The solid line is the concentration-mass fit proposed by \citet{Klypin:2014kpa} for the Planck cosmology, the shaded grey region around it representing a typical $1\sigma$ concentration scatter of 0.1 dex. }
    \label{fig:context}
\end{figure*}

Fig. \ref{fig:context} summarizes our results for the concentration-mass relation obtained from Lomonosov. In the same figure, we also put the Lomonosov simulation results into a more general context by comparing them against results from other simulations available at different halo mass scales \citep{2004ApJ...612...50C,2005Natur.433..389D,Diemand:2008in,Ishiyama:2011af,Anderhalden:2013wd,Sanchez-Conde:2013yxa,Ishiyama:2014uoa,2016MNRAS.457.3492H,Klypin:2014kpa}. The figure resembles a similar previous compilation by \citet{Sanchez-Conde:2013yxa}, this time including Lomonosov findings. More precisely, we depict \texttt{L512} concentration values and $1\sigma$-error bars (halo-to-halo variations), together with Lomonosov zoom regions concentrations obtained by applying the method described above to correct for environmental effects. We also include in the same figure halo concentration values found in other works from the smallest $\sim$10$^{-6}$ \msun~microhalos up to the largest galaxy-cluster-size structures with masses of about $\sim$10$^{15}$ \msun. 
Whenever needed, simulation results were scaled down to present time by multiplying their concentration values by $[H(z)/H(0)]^{2/3}$.  
We note that this differs from the most commonly adopted $(1+z)$ correction factor, also used in \citet{Sanchez-Conde:2013yxa}. 
The new correction factor more accurately accounts for the effect of the dark energy at late times, significantly reducing halo concentration with respect to that originally shown in Fig.1 of \citet{Sanchez-Conde:2013yxa} in some cases, and it has been derived within the framework of the model proposed by \citet{Maccio':2008xb} to evolve halo concentrations in a $\Lambda CDM$ universe. The latter is a refinement of the original model in \citet{Bullock:1999he}. 

\begin{table}
\caption{Median concentrations and 1-$\sigma$ scatters found in the zoom regions before and after correcting for  environmental effects.}
\centering
\begin{tabular}{ccccc}
\hline
\hline
Mass range, & \texttt{L4km} & \texttt{L4ko} & \texttt{L4ku} & corrected \\
\msun & & & & \\
\hline
$10^7 - 3\times10^7$  & $ 10.73_{-4.54}^{+5.43} $ & $ 10.72_{-4.75}^{+6.01} $ & $ 9.75_{-3.76}^{+3.94} $ & $ 10.44_{-4.36}^{+5.26} $ \\
$3\times10^7 - 10^8$ & $ 10.19_{-4.06}^{+4.89} $ & $ 10.27_{-4.29}^{+5.17} $ & $ 9.22_{-3.39}^{+3.21} $ & $ 9.89_{-3.89}^{+4.60} $ \\
$10^8 - 10^9$ & $ 9.22_{-3.62}^{+3.96} $ & $ 9.10_{-3.89}^{+4.16} $ & $ 8.42_{-2.92}^{+2.75} $ & $ 8.96_{-3.47}^{+3.76} $ \\
$10^9 - 10^{10}$ & $ 7.31_{-2.94}^{+2.77} $ & $ 7.52_{-3.12}^{+3.04} $ & $ 7.46_{-2.96}^{+2.54} $ & $ 7.39_{-3.00}^{+2.95} $ \\
\hline 
\hline
\end{tabular}
\label{tab:corr}
\end{table}

As already noted in \citet{Sanchez-Conde:2013yxa}, there is a lack of simulation results at intermediate halo masses, i.e. $\sim$10$^{-3}$ -- 10$^{7}$ \msun. We remind that filling this gap with new simulation data is one of our ultimate goals, and this work represents a first important step in this direction. Remarkably, Lomonosov provides the best halo statistics to date in the range $\sim$10$^{7}$ -- 10$^{9}$ \msun. Indeed, each of the Lomonosov concentration data points shown in Fig. \ref{fig:context} refer to a few hundred to several thousand halos,
 while the other few  points previously existing in this same mass range, also shown in the figure, were derived at most from only a few dozen halos each \citep{2004ApJ...612...50C,Ishiyama:2011af,Sanchez-Conde:2013yxa}.  
 In the high-mass end of our Lomonosov simulations, i.e. above 10$^{9}$ \msun, the halo statistics is still good enough to allow for some meaningful overlap and direct comparison with results from the MultiDark suite  \citep{Klypin:2014kpa}. We also note that the halo-to-halo scatter of Lomonosov concentrations is of the same order of the one found in previous works, of about 0.10 dex. 

It is remarkable the good agreement among the different simulation data sets within the involved uncertainties.
We also confirm, once again, the excellent agreement of simulation data with the semi-analytical $c(M)$ model of \citet{Prada:2011jf}, initially calibrated for the WMAP7 cosmology and then recently updated to the Planck cosmology in \citet{Klypin:2014kpa}. We recall that this $c(M)$ model is deeply rooted in the $\Lambda CDM$ cosmological framework itself by making a full correspondence between dark matter halo concentrations and the r.m.s. of matter fluctuations. 
We note that, in order to show the $c(M)$ relation given by this model all the way down to 10$^{-7}$ \msun, i.e. the minimum halo mass shown in Fig. \ref{fig:context}, we first computed the r.m.s. of matter fluctuations directly from the matter power spectrum that was used to generate the MultiDark simulations\footnote{But extrapolating it down to smaller halo masses with a simple power law, and placing an exponential mass cut-off at 10$^{-12}$ \msun, i.e., well below the range shown in Fig. \ref{fig:context}.} and, then, we used this r.m.s. of matter fluctuations to derive halo concentrations by adopting the relationship found between these two quantities in~\citet{Klypin:2014kpa} (their equation (25)).
The agreement between data and model is present at all simulated halo mass scales, including a new confirmation of the flattening of the c(M) relation at masses below $\sim$10$^{10}$ \msun. Indeed, we observe a clear departure from the simple power-law behaviour that has been traditionally reported at higher halo masses. Other $c(M)$ models have been recently proposed that would yield similar qualitative results as well, e.g., \citet{2014MNRAS.441..378L,2015MNRAS.452.1217C,2015ApJ...799..108D,2016MNRAS.460.1214L}.

\section{Conclusions}   \label{sec:conclusions}
In this work, we have introduced the new Lomonosov simulation suite consisting of one moderate resolution full box simulation, with box size 32 $h^{-1}$~Mpc, and three high resolution zoom-in re-simulations of overdense, underdense and mean density regions within the same box. The main purpose of the simulations is to allow for accurate measurements of dark matter halo concentrations at masses below those typically achievable in large cosmological simulations. We focus on the $10^{7}$ --- $10^{10}$ \msun~halo mass range. 

Achieving the high resolution that is required to resolve well low-mass halos results in a simulated volume that is much smaller than the typical volume needed to ensure Universe homogeneity. This fact may distort the halo median concentration values found in simulations, since concentration is known to depend on the local environment density (e.g. \citet{2017MNRAS.466.3834L}).
We solve the potential issue of measuring halo concentrations in small-volume high-resolution simulations by simulating the three mentioned different regions covering a large enough range of densities around the mean density of the Universe.

We confirm the dependency on local environment by making use of data not only from our Lomonosov simulations, but also from the Small MultiDark Planck \citep{Klypin:2014kpa} and Very Small MultiDark Planck simulations{\footnote{Available at www.cosmosim.org.}. 
Indeed, we find the concentration of low-mass halos to depend on the density of the environment (Fig. \ref{fig:cmd}), less concentrated halos inhabiting less dense regions and viceversa. Yet, we find its impact on concentration values to be small in comparison with the scatter of halo concentrations. We conclude that zoom simulations, despite their limited representativity of the volume of the Universe, can be effectively used for the measurement of halo concentrations, at least at the halo masses probed by our simulations.}

Despite the effect of environment on concentration being subdominant, we develop a novel technique to correct and extrapolate our concentration measurements in these zoom-in simulations to the whole volume with greater accuracy. The final Lomonosov concentration results at $z=0$ are shown in Fig. \ref{fig:context} and extend down to $10^{7}$ \msun~with good confidence and superb halo statistics. 
We find an excellent agreement of Lomonosov concentration values with the $c(M)$ model proposed by \citet{Prada:2011jf}, later updated by \citet{Klypin:2014kpa} for the Planck cosmology. Indeed, the combination of Lomonosov data with, very especially, data from the 
MultiDark simulations at larger halo masses allows us to confirm once
again the flattening in the c(M) relation that was expected to occur below $\sim10^{10}$ \msun~\citep{Sanchez-Conde:2013yxa}.

This work represents a first important step to measure halo concentrations at intermediate, yet vastly unexplored halo mass scales, down to the smallest ones.  Now, we possess a powerful, fully developed and tested technique to reliable extrapolate halo concentrations from small volumes up to meaningful ones. In the future, we plan to apply this technique to new high-resolution simulations in order to measure halo concentrations at even smaller scales than those probed in this work. Note that by learning about the key structural properties of low-mass dark matter halos we will be also testing fundamental $\Lambda$CDM predictions for these objects in greater detail. Also, our findings will be particularly relevant for the dark matter search community as it will allow for a more precise calculation of dark matter particle annihilation fluxes (e.g.~\citet{2015JETP..121.1104C,2017MNRAS.tmp...47M}).

All Lomonosov data and files are now public for community use and can be easily accessed via the \href{http://projects.ift.uam-csic.es/skies-universes/SUwebsite/index.html}{\it Skies and Universes} website.

\section*{Acknowledgements}
This paper is dedicated to the memory of our friend and colleague Dr. Pedro Colin, who first studied the properties of dwarf halos.

We thank Peter Behroozi for his help in tuning the \texttt{Rockstar} halo finder for our needs. 
We also thank \'Angeles Molin\'e for providing comments on a previous draft.
The work of SVP was supported by RFBR grant 16-32-00263. SVP is very grateful to the Instituto de F\'isica Te\'orica in Madrid, where part of this work was done. 
MASC is supported by the {\it Atracci\'on de Talento} contract no. 2016-T1/TIC-1542 granted by the Comunidad de Madrid in Spain, 
and also partially supported by MINECO under grant FPA2015-65929-P (MINECO/FEDER, UE).
MASC also acknowledges the support of the Swedish Wenner-Gren Foundations to develop part of this research. 
The authors also acknowledge the support from the MULTIDARK project of Spanish MCINN Consolider-Ingenio: CSD2009-00064, and the support of the Spanish MINECO's ``Centro de Excelencia Severo Ochoa'' Programme under grant SEV-2012-0249. MASC, FP acknowledge support from MINECO grant AYA2014-60641-C2-1-P. GY acknowledges support from  MINECO/FEDER under research grant AYA2015-63810-P.

The study was supported by the Supercomputing Center of Lomonosov Moscow State University.



\bibliographystyle{mnras}
\bibliography{halo} 


\appendix

\section{Rockstar halo finder set of parameters }
\label{ax:rockstar}
In order to compare our results with those of \citet{Klypin:2014kpa}, we follow their approach to measure halo concentrations as close as possible. We use version \texttt{Rockstar-0.99.9-RC3} and choose the following parameters:

\begin{lstlisting}
STRICT_SO_MASSES = 1
BOUND_PROPS = 0
MASS_DEFINITION = "200c"
\end{lstlisting}
Here \verb|STRICT_SO_MASSES=1| implies calculating a spherical overdensity in the full particle distribution, not among the FoF-selected particles, as it is done by default. However, we note that this spherical overdensity calculation is used only for the computation of halo masses, not for the derivation of $V_{max}$ which is the quantity needed to measure halo concentrations. Thus, we insert the calculation of $V_{max}$ into the file \verb|io/io_bgc2.c|.

\section{Validation tests of concentration extrapolations}
\label{ax:extrapol}

\begin{figure}
    \centering
    \includegraphics[width=\linewidth]{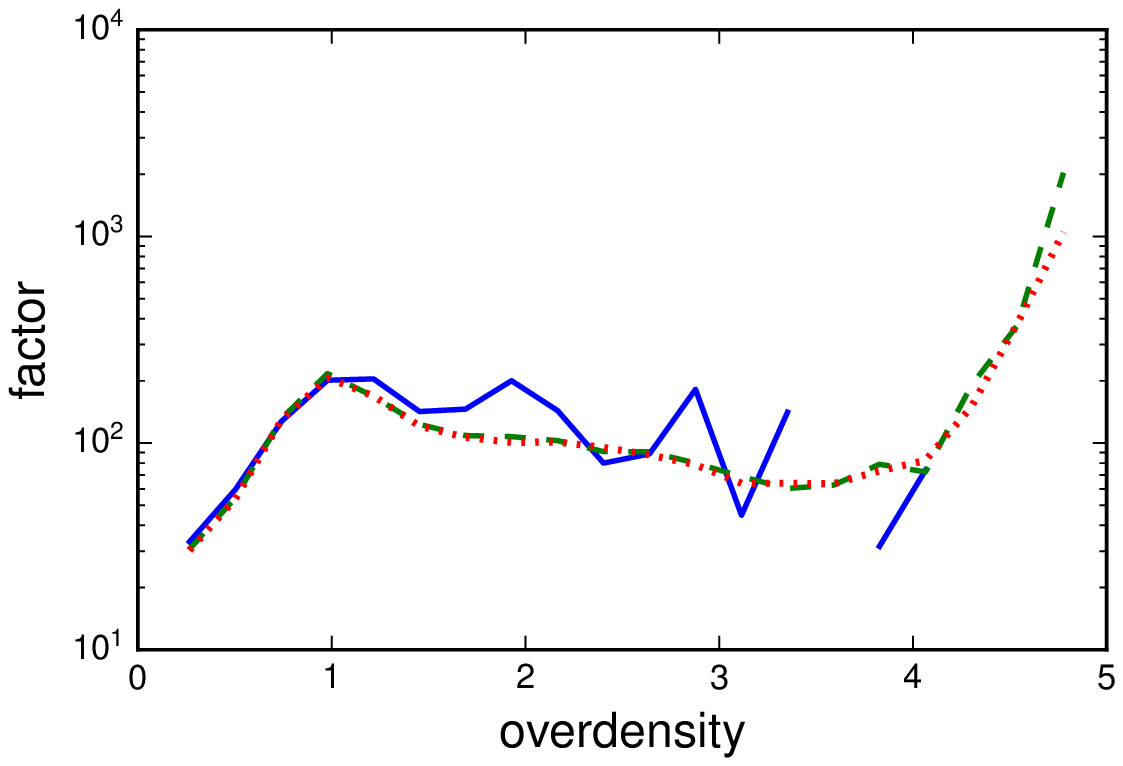}
    \caption{Volume correction factor for a uniform grid size of $30^3$ points (blue solid), $100^3$ (green dashed) and $200^3$ (red dotted).}
    \label{fig:test_factor}
\end{figure}

\begin{figure}
    \centering
    \includegraphics[width=\linewidth]{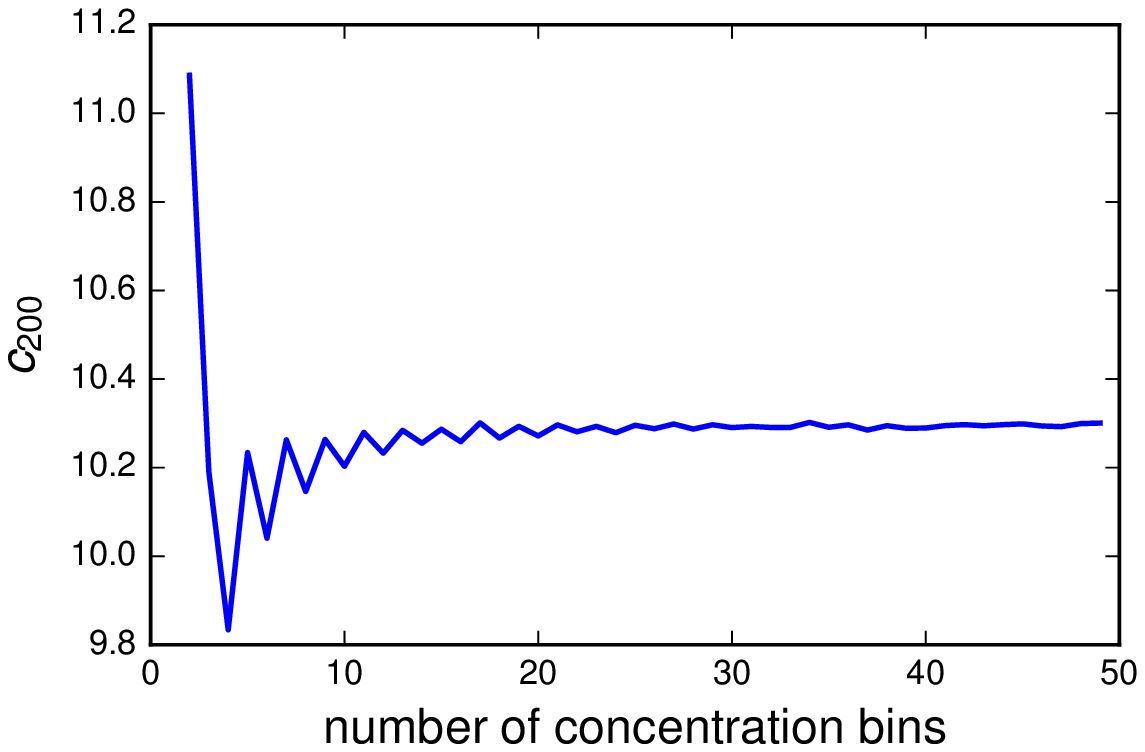}
    \caption{Extrapolated concentration values in the $10^7-10^8$ \msun~halo mass range as a function of number of bins in concentration.}
    \label{fig:test_nbins_c}
\end{figure}

\begin{figure}
    \centering
    \includegraphics[width=\linewidth]{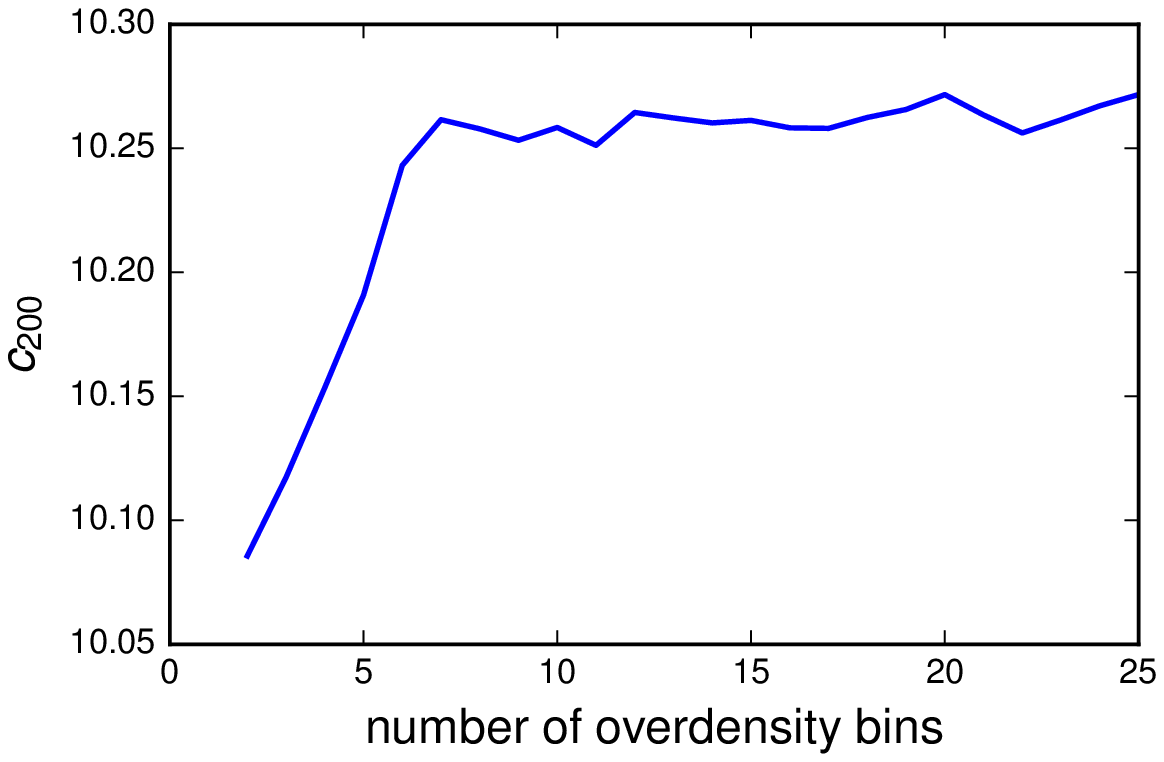}
    \caption{Extrapolated concentration values in the $10^7-10^8$ \msun~halo mass range as a function of number of bins in overdensity. }
    \label{fig:test_nbins_dens}
\end{figure}

First, we check for convergence of the measured concentration values by selecting optimal bin sizes and uniform grid size. We find that we need of the order of 100 cells in each spatial dimension, see Fig. \ref{fig:test_factor}.
As for the number of bins needed to properly build 2D histograms of overdensity and concentration, like those of Fig. \ref{fig:cmd} in the main text, it can be seen from Figs. \ref{fig:test_nbins_c}, \ref{fig:test_nbins_dens} that $\sim$20 bins for both concentration and overdensity are already enough.

\begin{figure}
    \centering
    \includegraphics[width=\linewidth]{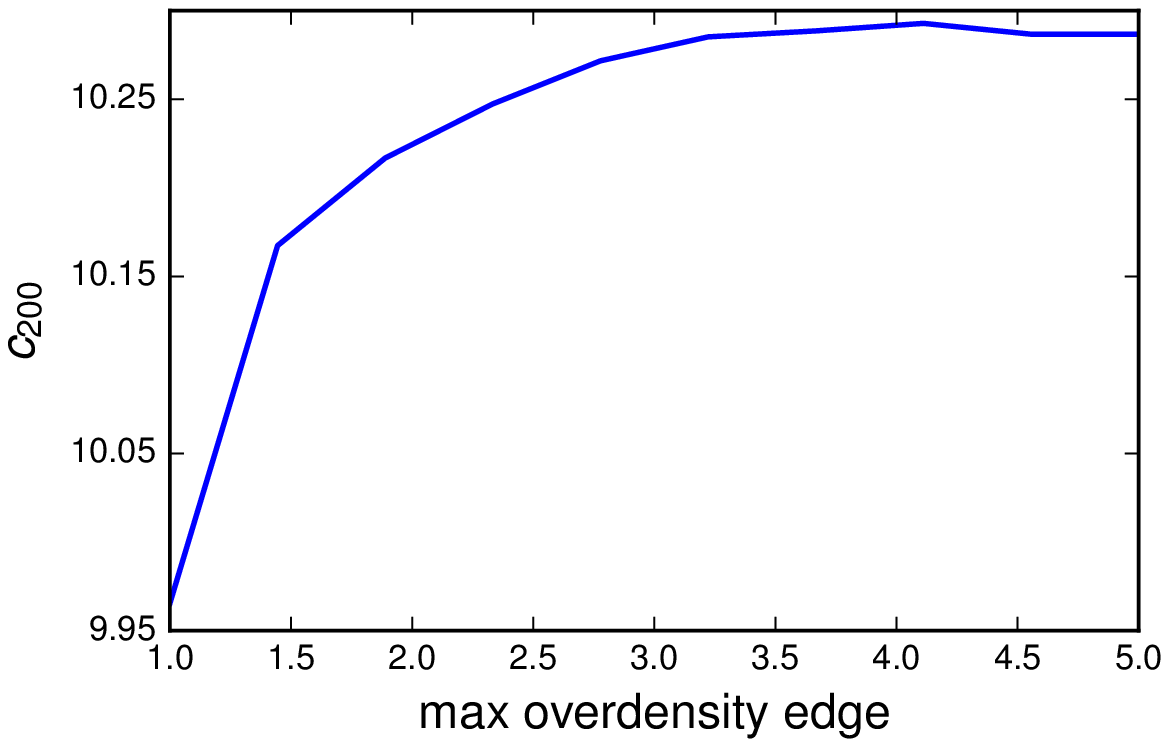}
    \caption{Extrapolated concentration values in the $10^7-10^8$ \msun~halo mass range as a function of maximum overdensity in the zoom regions. }
    \label{fig:test_dmax}
\end{figure}

\begin{figure}
    \centering
    \includegraphics[width=\linewidth]{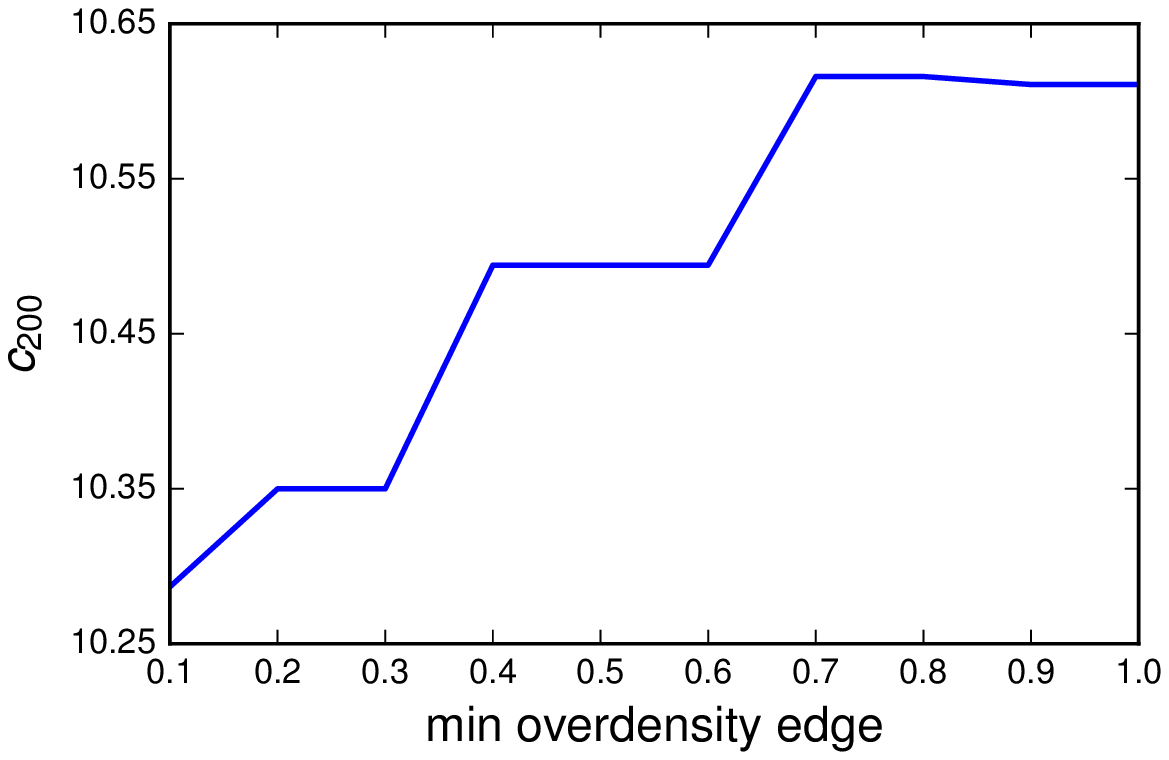}
    \caption{Extrapolated concentration values in the $10^7-10^8$ \msun~halo mass range as a function of minimum overdensity in the zoom regions.}
    \label{fig:test_dmin}
\end{figure}

We noticed that the full box simulation has cells reaching overdensities up to $\delta>10$ while in our zoom regions this upper value is limited to $\delta\sim5$. Since the correction factor is observed to grow as the overdensity increases, it may be possible that the high-density regions that are missing in the zoom simulations could be actually important in the derivation of the actual value of the correction factor. To check this, we vary the upper overdensity edge in our extrapolation method and find that it has a very small impact, see Fig. \ref{fig:test_dmax}; thus no additional correction is needed here. The impact of varying the lower overdensity edge seems to be more relevant, see Fig. \ref{fig:test_dmin}, yet the coverage of low overdensity values in the zoom simulations is almost full. This means that the diversity of different kinds of environment is sufficient in our zoom simulations for a correct extrapolation of halo concentrations to the full box.  

Finally, we check our method by comparing the extrapolated median concentration to that of the full box in the overlapping mass range where the \texttt{L512} simulation still has enough resolution and the zoom simulations still yield a decent statistics (110 halos), i.e.  $5\times10^9$ -- $10^{10}$ \msun. The extrapolated value we get in this mass range is 6.95, very close to the value of 6.90 we measure in the full box (even too good given the low halo statistics).

\begin{figure}
    \centering
    \includegraphics[width=\linewidth]{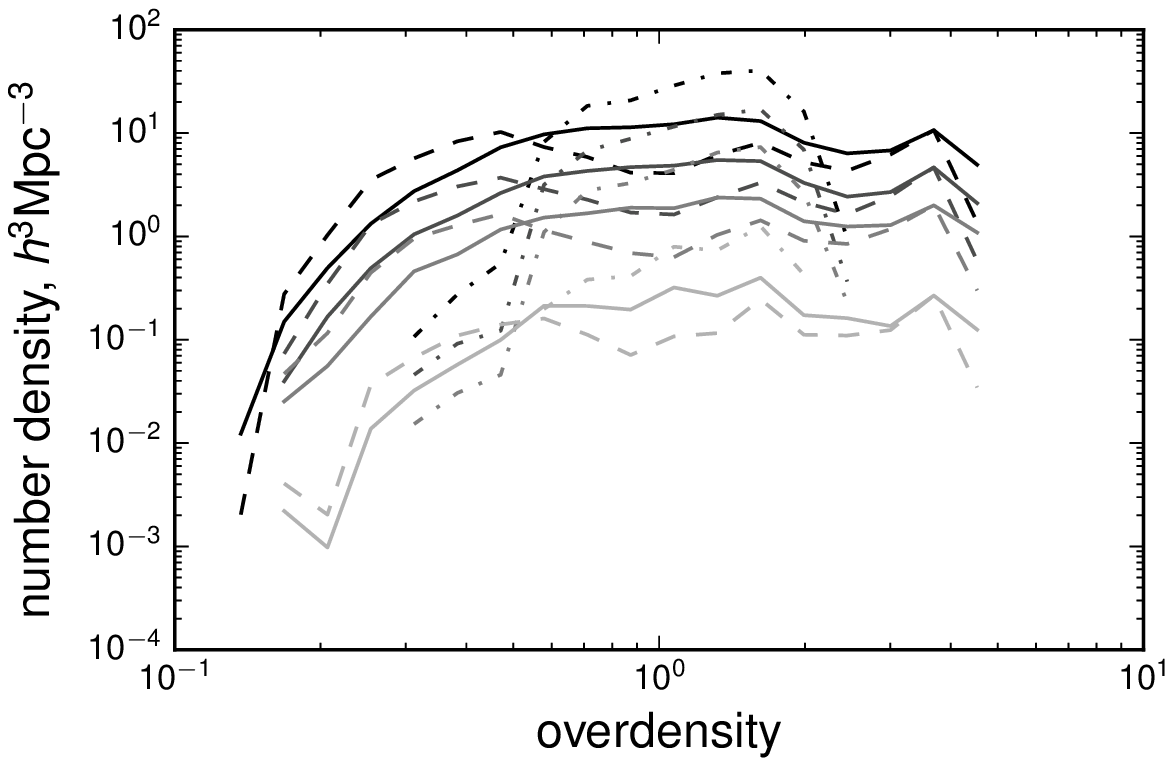}
    \caption{Distribution of halo overdensity for four mass bins, i.e., $[10^7,3\times10^7,10^8,10^9,10^{10}]$ \msun. From darker to lighter gray tones, lines correspond to larger and larger halo masses. Solid lines represent the volume-corrected distributions; dashed lines are those found when the three zoom regions are used together; dot-dashed are from the mean overdensity zoom region only (\texttt{L4km} simulation).}
    \label{fig:test_hist}
\end{figure}

We show partial results of the volume correction procedure in Fig. \ref{fig:test_hist}. In this Figure, histograms of overdensities are shown for the four considered mass bins before and after applying the correction. For comparison, overdensity histograms from the mean overdensity zoom simulation are also shown.

\section{$V_{max}/V_{200}$ for low mass halos}
\label{ax:vmax}
In Table \ref{tab:vmax}, we provide the corrected median $V_{max}/V_{200}$ and 1$\sigma$ scatter for the four considered Lomonosov halo mass bins.

\begin{table}
\caption{Corrected median $V_{max}/V_{200}$ values and associated 1$\sigma$ scatter for Lomonosov halos in the full Lomonosov suite.}
\centering
\begin{tabular}{cc}
\hline
\hline
Mass range, & $V_{max}/V_{200}$  \\
\msun &  \\
\hline
$10^7 - 3\times10^7$ & $ 1.22_{-0.12}^{+0.13} $ \\
$3\times10^7 - 10^8$ & $ 1.20_{-0.11}^{+0.12} $ \\
$10^8 - 10^9$ & $ 1.18_{-0.10}^{+0.10} $ \\
$10^9 - 10^{10}$ & $ 1.13_{-0.09}^{+0.08} $ \\
\hline
\hline
\end{tabular}
\label{tab:vmax}
\end{table}


\bsp	
\label{lastpage}
\end{document}